\documentclass[10pt,conference]{IEEEtran}
\usepackage{cite}
\usepackage{amsmath,amssymb,amsfonts}
\usepackage{graphicx}
\usepackage{textcomp}
\usepackage{xcolor}
\usepackage[hyphens]{url}
\usepackage{fancyhdr}
\usepackage[hidelinks]{hyperref}
\usepackage{algorithm}
\usepackage{algpseudocode}
\usepackage{subcaption}
\usepackage{multirow}
\usepackage{tabularx}
\usepackage{array}

\newcommand{\hpcayear}{2027}

\newcommand{\revv}[1]{\textcolor{black}{#1}}
\newlength{\responsevspace}
\title{GPS: \underline{G}uiding \underline{P}hase-Level Prefetcher \underline{S}election with \\ Data-Driven Method}

\def\hpcacameraready{}

\newcommand\hpcaauthors{
    Buqing Xu\textsuperscript{\textdagger},
    Mengming Li\textsuperscript{\textdagger},
    Qijun Zhang,
    Ziyan Guo,
    and Zhiyao Xie\textsuperscript{*}
}

\newcommand\hpcaaffiliation{
    The Hong Kong University of Science and Technology
}

\newcommand\hpcaemail{
    \{bxuax, mengming.li, qzhangcs, zguoby\}@connect.ust.hk,
    eezhiyao@ust.hk
}

\author{
  \ifdefined\hpcacameraready
    \IEEEauthorblockN{\hpcaauthors{}}
      \IEEEauthorblockA{
        \hpcaaffiliation{} \\
        \hpcaemail{}
      }
  \else
    \IEEEauthorblockN{\normalsize{HPCA \hpcayear{} Submission
      \textbf{\#612}} - Confidential Draft - Do NOT Distribute!!
    }
  \fi 
}

\fancypagestyle{camerareadyfirstpage}{%
  \fancyhead{}
  
  \fancyhead[C]{
    \ifdefined\aeopen
    \parbox[][12mm][t]{13.5cm}{\hpcayear{} IEEE International Symposium on High-Performance Computer Architecture (HPCA)}    
    \else
      \ifdefined\aereviewed
      \parbox[][12mm][t]{13.5cm}{\hpcayear{} IEEE International Symposium on High-Performance Computer Architecture (HPCA)}
      \else
      \ifdefined\aereproduced
      \parbox[][12mm][t]{13.5cm}{\hpcayear{} IEEE International Symposium on High-Performance Computer Architecture (HPCA)}
      \else
      \parbox[][0mm][t]{13.5cm}{\hpcayear{} IEEE International Symposium on High-Performance Computer Architecture (HPCA)}
    \fi 
    \fi 
    \fi 
    \ifdefined\aeopen 
      \includegraphics[width=12mm,height=12mm]{ae-badges/open-research-objects.pdf}
    \fi 
    \ifdefined\aereviewed
      \includegraphics[width=12mm,height=12mm]{ae-badges/research-objects-reviewed.pdf}
    \fi 
    \ifdefined\aereproduced
      \includegraphics[width=12mm,height=12mm]{ae-badges/results-reproduced.pdf}
    \fi
  }
  \fancyfoot[C]{}
}
\begin{document}
\maketitle

\ifdefined\hpcacameraready 
  \thispagestyle{camerareadyfirstpage}
  \pagestyle{empty}
\else
  \thispagestyle{plain}
  \pagestyle{plain}
\fi

\newcommand{\hpcaheight}{0mm}
\ifdefined\eaopen
\renewcommand{\hpcaheight}{12mm}
\fi

{
\renewcommand{\thefootnote}{\fnsymbol{footnote}}
\footnotetext[2]{These authors contributed equally.}
\footnotetext[1]{Corresponding author.}
}

\begin{abstract}
Hardware prefetching is a widely used technique for hiding off-chip memory latency. Modern processors increasingly adopt composite prefetchers that combine multiple prefetching engines to handle the diverse memory access patterns found in real applications. However, coordinating multiple prefetchers requires a policy selection algorithm that maps observed system feedback to appropriate prefetching policies. Existing approaches suffer from two key limitations. First, relying on only a small set of features can lead to misleading prefetcher policies due to incomplete visibility into the system state. Second, limited on-chip hardware resources force these designs to use simple mapping mechanisms that cannot effectively navigate the large prefetching policy space.\looseness=-1

To address these challenges, we propose \textbf{GPS}, a data-driven framework for prefetcher selection that leverages comprehensive profiling and offline learning. GPS collects a rich set of features that capture both system behaviors and hardware resource states, and uses an offline machine learning model to learn the mapping between these features and the IPC-best policy.
By offloading complex policy generation to software, GPS eliminates the need for costly online exploration and enables lightweight runtime execution through pre-generated policy hints.
GPS reduces the performance gap to only 2.6\%, compared to 13.6\% and 14.2\% in prior state-of-the-art methods.
We evaluate GPS on SPEC CPU2006, SPEC CPU2017, and GAP benchmark suites. 
GPS outperforms the state-of-the-art prefetcher selection method I-POP by 12.14\%, 8.34\%, and 14.17\% in geometric-mean IPC, respectively. GPS achieves these gains with only 778 bits of additional on-chip storage for runtime policy control.\looseness=-1
\end{abstract}

\section{Introduction}
Data prefetching is widely used to mitigate the CPU memory wall by predicting and fetching data before it is requested by the processor~\cite{vanderwiel2000data}. Its effectiveness therefore depends critically on the ability to accurately predict future memory accesses. However, memory access patterns vary substantially across applications and may also change across execution stages within the same application. To accommodate such diversity, researchers have developed specialized hardware prefetchers for different classes of access patterns, including stream~\cite{he_dsdp_2022, hur2006memory, shakerinava2019multi}, stride~\cite{jouppi_improving_1990, baer_effective_1991, dahlgren_effectiveness_1995, kim_stride-directed_1997, fu1992stride}, spatial~\cite{bakhshalipour_bingo_2019, bera_dspatch_2019, ishii_access_2009, jiang_merging_2022, somogyi_spatial_2006, lin2024pars, kim_path_2016, michaud_best-offset_2016, pugsley_sandbox_2014, shevgoor2015efficiently}, and temporal prefetchers~\cite{bakhshalipour_domino_2018, jain_linearizing_2013, nesbit2004data, wenisch_practical_2009, wu_triage_2019, wu2019efficient, joseph1997prefetching, ainsworth_triangel_2024, jiang2026thoth,li2026icp}. Because no single prefetcher performs well across all workloads and execution stages, modern processors increasingly combine multiple complementary prefetchers into a composite prefetching system~\cite{pakalapati_bouquet_2020, intel_atom_prefetch}.

The integration of multiple prefetchers, however, introduces a fundamental coordination challenge: the system must determine which prefetchers should handle the current memory access stream. An inappropriate selection may generate inaccurate or redundant prefetches, wasting cache capacity, request-buffer entries, and memory bandwidth, and may ultimately degrade performance~\cite{li2025integrating}. We formulate the prefetcher selection problem as mapping input features (e.g., runtime information used to guide selection) to a prefetching policy:  \emph{$f(\text{features}) = \text{policy}$}.
The policy includes how demand requests are allocated, which prefetcher can generate requests, and what prefetch degree is assigned.
Two key components determine the effectiveness of a prefetcher selection method: (1) the \textbf{input features} it collects, ranging from basic hardware counters to derived features such as prefetch accuracy, and (2) the mapping algorithm \emph{$f$}, which \textbf {maps} these features to a policy. However, we observe that existing prefetcher selection methods~\cite{kondguli_division_2018, pakalapati_bouquet_2020, gerogiannis_micro-armed_2023, li2025integrating, lin2026pop} remain limited in both aspects: (1) they often rely on incomplete input features and (2) they typically employ heuristic or static mapping algorithms that are too rigid to handle diverse workload behaviors.
These limitations prevent existing prefetcher selection methods from using the IPC-best policy (the policy that achieves the highest instructions per cycle (IPC) among all prefetcher policies evaluated in our experiments).
We further elaborate these limitations below. 

\begin{figure}[t]
    \centering
    \includegraphics[width=1\linewidth]{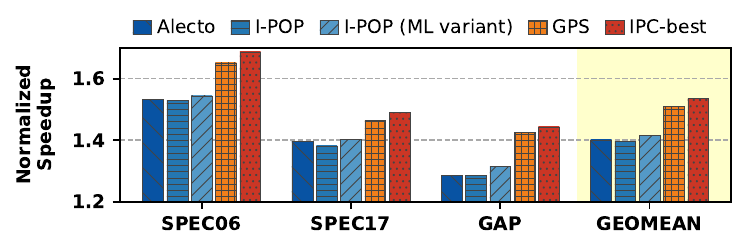}
    \vspace{-.2in}
    \caption{GPS achieves speedup close to the IPC-best baseline and consistently outperforms state-of-the-art methods.}
    \vspace{-.2in}
    \label{fig:speedup_best}
\end{figure}

\textbf{Limitation 1: Incomplete input features.} Selecting the best prefetcher requires a global view of the overall system status. For example, when memory bandwidth utilization approaches a critical threshold, the selection policy should become more conservative to prioritize demand accesses. Conversely, under low memory pressure, even an aggressive policy with relatively low accuracy can still improve performance by increasing coverage. Therefore, approaches that rely on limited input features (e.g., Alecto~\cite{li2025integrating}, which depends solely on prefetch accuracy)
can be misleading, as they ignore other important system states (e.g., memory bandwidth).
I-POP~\cite{lin2026pop} further adopts approximately 10 input features, considering cache hits, cache pollution, and basic resource contention.
However, it still lacks comprehensive visibility into other system states, such as the Miss Status Holding Register (MSHR) state.
Incorporating the MSHR state can provide insights into the true effectiveness of timely prefetches. Consequently, such incomplete input features hinder the identification of the IPC-best policy.

\textbf{Limitation 2: Oversimplified mapping from input features to a prefetching policy.}
Existing methods~\cite{kondguli_division_2018, pakalapati_bouquet_2020, gerogiannis_micro-armed_2023, li2025integrating, lin2026pop} rely on heuristic or static mapping algorithms, which fail to capture the nonlinear relationship between input features and prefetching policies. For example, I-POP~\cite{lin2026pop} uses a Prefetch Effectiveness (PE) score with a linear analytical formulation to map input features to a policy. We observe that such mappings can lead to suboptimal performance. To demonstrate this limitation, we evaluate a variant named \textit{I-POP (ML variant)}, which replaces the linear formulation in I-POP with a nonlinear machine learning (ML) model (i.e., a multilayer perceptron (MLP)~\cite{goodfellow2016deep}) while keeping the same input features.
As shown in Figure~\ref{fig:speedup_best}, this direct substitution improves performance by 2.2\% over the original I-POP.
This result implies that the linear model is insufficient for predicting the IPC-best policy using the features employed by I-POP. Furthermore, if we introduce the richer input features required by Limitation 1, the mapping problem becomes even more complex, making simple mapping algorithms increasingly inadequate.\looseness=-1

\textbf{Our Solution.} To address these two limitations, we propose \textbf{GPS}.
GPS moves beyond limited runtime features and simple mapping algorithms by using detailed system profiling and an ML model.
GPS operates at phase granularity: a phase is an interval of one million retired instructions and serves as the basic unit for profiling and policy updates.
For each phase, we identify 55 features that help predict the IPC-best policy.
Specifically, these features are detailed in two aspects. First, they capture diverse system behaviors, such as cache behavior (hits, misses, and latency), MSHR behavior (hits, misses, and latency), prefetch metrics, branch metrics, etc. Second, they span a wide range of hardware structures, including multiple cache levels (L1, L2, L3), prefetchers, MSHRs, and branch predictors. To effectively utilize these features, we adopt a lightweight ML model to learn the mapping between features and an IPC-best policy based on known data.\looseness=-1

Our solution, GPS, adopts a profile-guided method that offloads the hardware-unfriendly process---model training and inference---to software.
\textbf{Offline training}: The model is trained using \emph{system features} and the normalized IPC scores. After training, the model can predict IPC-best policies for new unseen workloads during inference.
\textbf{Offline inference}: First, the unknown target workload is profiled to generate phase-level hardware features. The ML model then processes these features to generate a sequence of policy hints, specifying the suitable prefetcher actions for each phase.
\textbf{Runtime usage}: At runtime, the system simply reads these pre-generated hints to guide the runtime prefetcher selection. Because these hints are phase-level, GPS can adapt its policy selection to each execution phase of the program.

\textbf{Novelty.}
GPS is a data-driven prefetcher selection framework that combines complete program profiling with an offline ML model.
Unlike existing works~\cite{gerogiannis_micro-armed_2023, li2025integrating, lin2026pop} that rely on limited runtime features and heuristic decision logic, GPS systematically collects rich phase-level system features and directly learns the mapping between these features and the IPC-best policy.
This solution changes prefetcher selection from a reactive, heuristic-driven process into a profile-guided, data-driven framework.
As a result, GPS not only addresses the limitations of incomplete input features and inaccurate mapping algorithms in prior work, but also provides a new perspective for achieving IPC-best policy selection with low online hardware complexity.

\textbf{Performance.} To quantify the benefit of our method, we compare existing prefetcher selection methods against an IPC-best baseline\footnote{IPC-best baseline uses the policy that yields the highest IPC in each phase. We derive the IPC-best baseline by exhaustively evaluating 279 candidate policies per phase and selecting the one maximizing IPC for each phase.}.
As shown in Figure~\ref{fig:speedup_best}, Alecto~\cite{li2025integrating} and I-POP~\cite{lin2026pop} underperform the IPC-best baseline by 13.6\% and 14.2\%, respectively, highlighting the limitations of incomplete input features and rigid mapping algorithms in existing methods. In contrast, GPS largely overcomes these limitations and reduces the performance gap to only 2.6\%.

We summarize our contributions below:
\begin{itemize}
    \item We conduct a thorough analysis of state-of-the-art prefetcher selection methods to uncover their key limitations. By comparing the policies selected by existing methods with those selected by GPS and the IPC-best baseline, we clearly expose the shortcomings of existing methods and provide a detailed analysis of why current mechanisms fail to achieve IPC-best prefetcher selection.
    \item We address these limitations with GPS, a data-driven framework for prefetcher selection. Unlike prior work, which relies on a small set of runtime features and a rigid mapping algorithm, GPS collects 55 features and uses an offline ML model to predict the IPC-best policy for each phase, reducing the gap to the best performance from 13.6\%--14.2\% to only 2.6\%.
    \item We evaluate GPS on SPEC CPU2006~\cite{henning2006spec}, SPEC CPU2017~\cite{spec2017}, and GAP~\cite{beamer2015gap} benchmark. GPS outperforms the state-of-the-art prefetcher selection method I-POP~\cite{lin2026pop} by 12.14\%, 8.34\%, and 14.17\% in geometric-mean IPC, respectively.
\end{itemize}
\section{Background and Motivation}
\subsection{Composite Hardware Prefetchers}
No single prefetching method consistently performs well across all complex workloads because real applications often exhibit multiple access patterns simultaneously.
Thus, composite prefetchers~\cite {pakalapati_bouquet_2020, kondguli_division_2018, gerogiannis_micro-armed_2023, li2025integrating, lin2026pop}, which use multiple prefetching algorithms, show great potential in complex real-world workloads.
However, integrating multiple prefetchers introduces the challenge of coordinating them, which we refer to as the prefetcher selection problem.
Current prefetcher selection algorithms mainly make the following decisions~\cite{li2025integrating}:
\begin{itemize}
    \item \textbf{Demand request allocation} decides which prefetcher receives incoming demand requests for training.
    \item \textbf{Prefetch request filtering} decides which prefetchers are allowed to generate prefetch requests.
    \item \textbf{Prefetch degree adjustment} decides the aggressiveness of prefetchers.
\end{itemize}

Prefetcher selection methods determine prefetching policies.
We classify selection methods as \textit{static} or \textit{dynamic} based on whether they can adjust the generated policy during program execution.\looseness=-1

\subsubsection{Static Prefetcher Selection Methods}
Static methods apply a fixed policy throughout execution.
A representative example of this category is IPCP~\cite{pakalapati_bouquet_2020}.
IPCP uses a fixed, rule-based policy to coordinate its multiple lightweight prefetchers. It assigns each prefetcher a fixed priority, filters out prefetch requests from lower-priority prefetchers, and assigns a static prefetch degree to each prefetcher.
DOL~\cite{kondguli_division_2018} uses static priority arbitration to allocate each demand request to the highest-priority prefetcher and assigns each prefetcher a fixed prefetch degree.
Although these static prefetcher selection methods simplify hardware implementation, they lack flexibility across diverse workloads, potentially diminishing prefetching effectiveness.
For instance, in program phases containing access patterns that are better suited to a lower priority prefetcher, the static priority logic may lead to suboptimal policy selection, as it cannot adapt the priority order to the workload's dynamic requirements.\looseness=-1

\begin{figure}
    \centering
    \includegraphics[width=1\linewidth]{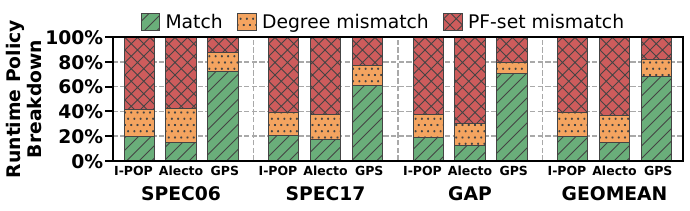}
    \vspace{-.2in}
    \caption{\revv{Runtime policy selection breakdown for I-POP~\cite{lin2026pop}, Alecto~\cite{li2025integrating}, and GPS. Bars show the proportions of Match, Degree mismatch, and PF-set mismatch against the top 10\% policies ranked by IPC. GPS achieves the highest match rate across all three benchmark suites.}}
    \vspace{-.2in}
    \label{fig:policy_breakdown}
\end{figure}

\subsubsection{Dynamic Prefetcher Selection Methods}
Dynamic methods adjust prefetching policies based on observed program behavior.
Early works such as FDP~\cite{srinath2007feedback} and HPAC~\cite{ebrahimi2009coordinated} use predefined heuristic rules to control prefetch aggressiveness.
Bandit~\cite{gerogiannis_micro-armed_2023} employs reinforcement learning to dynamically adjust its policy, but its control scope is highly restricted. Among the three core decisions, it focuses solely on prefetch degree adjustment.
Crucially, its generated policy completely lacks demand request allocation and prefetch request filtering.
Because Bandit applies a single global degree adjustment to the entire demand access stream at a coarse time granularity, it cannot precisely route different requests to specific prefetchers.
This narrow control mechanism severely reduces its effectiveness when memory access patterns are interleaved.

Alecto~\cite{li2025integrating} manages policies at a finer PC level, using runtime accuracy for demand request allocation (which indirectly achieves prefetch request filtering) and prefetch degree adjustment. However, Alecto is severely limited by both its input features and its mapping algorithm.
First, relying only on prefetch accuracy as a feature can be misleading.
Second, Alecto employs a simple state machine to make policy selections.
This simple algorithm lacks the capacity to process dynamic system behaviors and navigate the large policy search space effectively.
We will further demonstrate this misleading issue in Section~\ref{Limitations_of_current_pf}.

\begin{figure}[t]
    \centering
    \begin{subfigure}{0.49\linewidth}
        \centering
        \includegraphics[width=\linewidth]{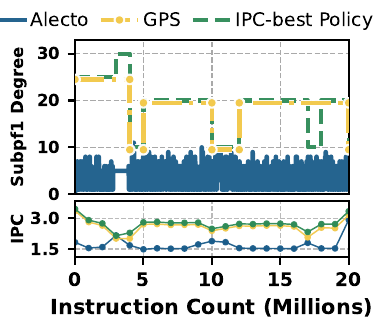}
        \caption{Alecto converges to a suboptimal policy.}
        \label{fig:feature_misleading:a}
    \end{subfigure}\hfill%
    \begin{subfigure}{0.49\linewidth}
        \centering
        \includegraphics[width=\linewidth]{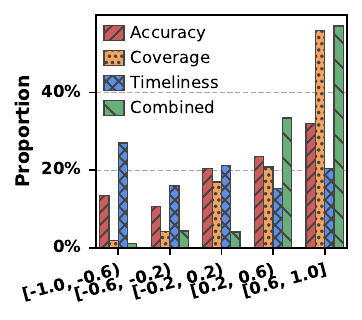}
        \caption{$\rho$ distribution between phase-level IPC and features.}
        \label{fig:feature_misleading:b}
    \end{subfigure}
    \caption{Feature misleading example. (a) Prefetch degree adjustment of a prefetcher and phase-level IPC (normalized to no prefetching) across Alecto, GPS, and the IPC-best policy in workload \textit{milc}. (b) Correlation distribution between phase-level IPC and runtime features, evaluated at a 1M-instruction granularity across workloads in SPEC CPU2006~\cite{henning2006spec}.}
    \vspace{-.15in}
    \label{fig:feature_misleading}
\end{figure}

I-POP~\cite{lin2026pop} introduces prefetch effectiveness (PE) in features to dynamically select the prefetching policy.
However, it still suffers from two key limitations.
First, its PE calculation is overly idealistic and fails to reflect true performance gains, particularly in memory-intensive workloads, causing severe feature misleading that prevents it from finding the IPC-best policy.
Second, I-POP still relies on a simple state machine to map these features to policy selections. This simplistic algorithm lacks the capacity to navigate the massive policy search space.
Furthermore, in complex workloads, all the above reactive prefetcher selection frameworks
~\cite{gerogiannis_micro-armed_2023, li2025integrating, lin2026pop} require substantial time to converge, inevitably causing a response lag.\looseness=-1

\subsection{Limitations of Current Prefetcher Selection Methods}\label{Limitations_of_current_pf}
By evaluating policies collected on the SPEC06, SPEC17, and GAP benchmarks (configuration is detailed in Section~\ref{sec:exep_method}),
\revv{we first analyze the behavior of state-of-the-art dynamic prefetcher selectors~\cite{li2025integrating, lin2026pop} across three benchmark suites by comparing their runtime policies with the IPC-best policy. We then identify three critical limitations in these dynamic prefetcher selectors.}

\revv{\textbf{Ineffectiveness of Current Prefetcher Selection Methods.}}
\revv{Figure~\ref{fig:policy_breakdown} shows the runtime policy breakdown for I-POP~\cite{lin2026pop}, Alecto~\cite{li2025integrating}, and GPS.
We classify the runtime policies into three categories: \emph{Match}, \emph{Degree mismatch}, and \emph{PF-set mismatch}, based on comparison with the IPC-best policy.
\emph{Match} means that the selected policy uses the same prefetcher set as the IPC-best policy, with each degree within 20\% of its corresponding reference value. \emph{Degree mismatch} means the set is the same, but at least one degree falls outside this range. \emph{PF-set mismatch} means the sets differ. For example, if the IPC-best policy is PF1 with degree 10, PF1 with degree 9 is a \emph{Match}, PF1 with degree 5 is a \emph{Degree mismatch}, and PF2 with degree 10 is a \emph{PF-set mismatch}. 
In practice, we use the top 10\% of evaluated policies ranked by IPC as the reference list. We assign \emph{Match} if the selected policy matches any reference. Otherwise, we assign \emph{Degree mismatch} if any reference has the same prefetcher set, and \emph{PF-set mismatch} if none does.
Figure~\ref{fig:policy_breakdown} shows that GPS achieves \emph{Match} rates of 61.0\% to 71.9\% across SPEC06, SPEC17, and GAP, compared with 19.1\% to 20.4\% for I-POP and 12.8\% to 17.4\% for Alecto.}

\begin{figure}
    \centering
    \includegraphics[width=\linewidth]{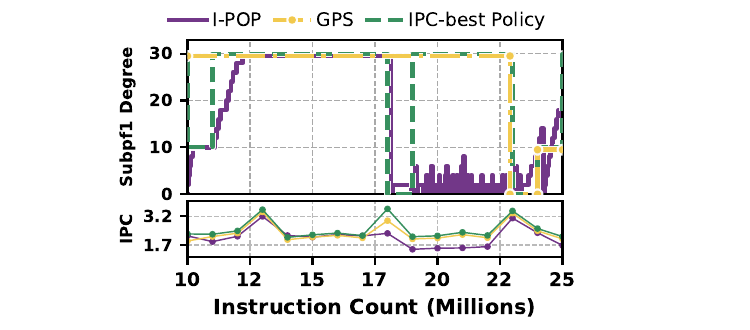}
    \vspace{-.2in}
    \caption{Simplistic mapping example. I-POP misses the IPC-best policy after 18 million instructions in workload \textit{leslie3d}.}
    \vspace{-.25in}
    \label{fig:simplistic_mapping}
\end{figure}

\textbf{Feature Misleading.}
An ideal prefetcher selection method identifies the prefetching policy that maximizes IPC.
\revv{Feature misleading occurs when runtime features do not always reflect actual IPC gains, leading to suboptimal policy selection.}
Existing dynamic prefetcher selection methods rely on one or more runtime features to predict suitable prefetching policies.
To illustrate this limitation, Figure~\ref{fig:feature_misleading:a} compares the dynamic degree of a sub-prefetcher under Alecto, GPS, and the IPC-best policy in the top panel, along with their phase-level IPC in the bottom panel.
Alecto relies solely on prefetch accuracy to determine the prefetching degree, where higher accuracy leads to a more aggressive prefetch degree.

As the two panels show, Alecto maintains a low prefetch degree, creating a massive IPC gap compared to the highly aggressive IPC-best policy.
In contrast, GPS successfully maintains an aggressive prefetch degree, achieving performance that significantly outperforms Alecto in the bottom panel.
This case highlights a critical finding: even when prefetch accuracy is low, the system can still benefit from a higher prefetch degree.
This occurs because sufficient memory bandwidth allows the system to tolerate lower accuracy, improving overall IPC by increasing prefetch coverage.

Figure~\ref{fig:feature_misleading:b} further presents the Spearman correlation coefficients ($\rho$) between phase-level IPC and various prefetching features under an evaluation setup similar to I-POP~\cite{lin2026pop}.
For each phase of one million instructions, we compute the $\rho$ between the actual IPC and the corresponding feature values across all prefetchers.
We evaluate three basic features (accuracy, coverage, and timeliness) alongside a newly derived combined feature that is explicitly engineered to maximize this correlation\footnote{After evaluating various nonlinear combinations of prefetch features, the best combination that maximizes $\rho$ is $\frac{\text{cov}^p}{1+w\cdot(1-\text{acc})^q}$.}.
As shown in Figure~\ref{fig:feature_misleading:b}, only about 50\% of the execution phases exhibit a strong positive correlation (0.6 to 1.0) for the basic features, while nearly 20\% even exhibit a negative correlation ($\rho < 0$). More importantly, even the optimized combined feature fails to guarantee a strong (only 57.14\%) positive correlation across all phases. This demonstrates that relying on simple prefetching features, or even their engineered combinations, can mislead policy selection.\looseness=-1

\begin{figure}
    \centering
    \includegraphics[width=1\linewidth]{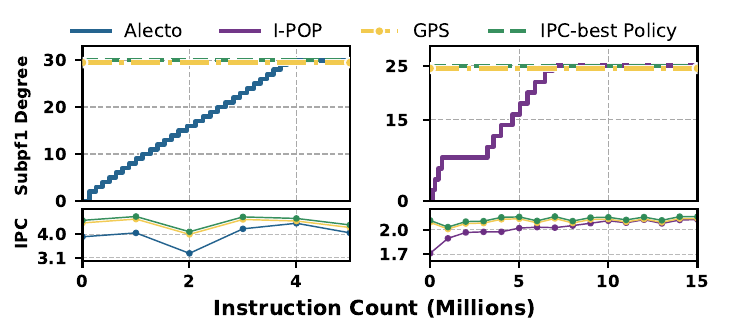}
    \vspace{-.2in}
    \caption{Response lag example. Alecto and I-POP require about four million and six million instructions to converge on \textit{soplex} and \textit{lbm}, respectively.}
    \label{fig:response_lag}
    \vspace{-.25in}
\end{figure}

\textbf{Simplistic Mapping.}
The reliance on heuristic rules or fixed formulations prevents mapping algorithms from effectively combining more features into decisions that maximize IPC.
For example, I-POP~\cite{lin2026pop} tracks up to 10 features, significantly more than Alecto. However, it still relies on a fixed formulation, failing to accurately capture the complex and nonlinear relationship between features and performance.
Figure~\ref{fig:simplistic_mapping} illustrates how this fixed formulation leads to suboptimal decisions.
Specifically, I-POP calculates a Prefetch Effectiveness (PE) score using a static equation, which subtracts cache pollution and DRAM contention penalties from prefetch hit benefits.
As shown in Figure~\ref{fig:simplistic_mapping}, I-POP successfully matches the IPC-best policy before 18 million instructions, benefiting from its richer set of tracked features.
However, this fixed calculation cannot adapt to dynamic system states. When memory accesses increase after 18 million instructions, I-POP heavily overestimates the negative impact of DRAM bus contention.
Consequently, it drops the prefetch degree and fails to discover an aggressive IPC-best policy.

\textbf{Response Lag.}
In addition, dynamic prefetcher selection methods suffer from response lag because they rely on accumulating sufficient runtime feedback to infer future program behavior and determine current prefetching policies.
To validate this limitation, Figure~\ref{fig:response_lag} compares the Alecto and I-POP runtime prefetching policies with the corresponding IPC-best policy.
It shows that, for certain workloads, Alecto and I-POP require approximately four and six million instructions to converge to their respective IPC-best policies.
We find that sparse but stable memory access patterns force these prefetchers to spend more time accumulating runtime feedback, causing them to miss the opportunity to apply the IPC-best policy in time.

\subsection{Motivation}
To tackle the limitations of prior works, our solution, GPS, is driven by the following motivations.
\subsubsection{Overcoming incomplete information using rich profiling}
A prefetcher selection algorithm requires a global view of the entire memory subsystem to make accurate decisions. 
Instead of using limited metrics, GPS collects 55 features during the profiling stage. 
These features span a wide range of hardware structures, including multiple cache levels, MSHRs, prefetchers, and branch predictors. 
From these structures, we gather raw hardware counters that capture diverse memory behaviors, such as hits, misses, and latency for both caches and MSHRs, alongside memory bandwidth and prefetch events. 
We then derive meaningful features from these counters, such as prefetch accuracy and cache miss rates. 
This comprehensive profiling provides sufficient context to identify the best policy across diverse system states, whether across different phases within a single workload or across different workloads.

\begin{figure}[t]
    \centering
    \includegraphics[width=0.85\linewidth]{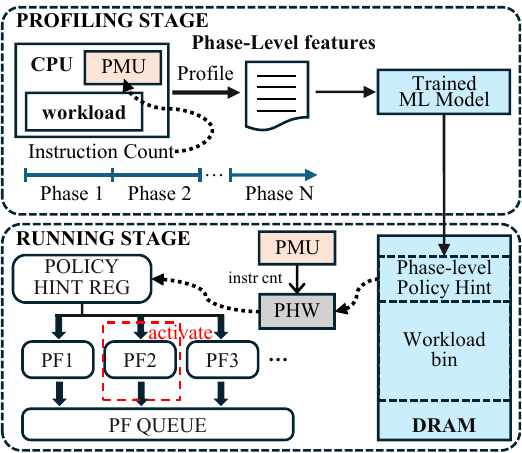}
    \vspace{-.05in}
    \caption{End-to-end GPS deployment workflow. During profiling, GPS uses a trained ML model to generate phase-level policy hints. During execution, a PHW, triggered directly by the PMU, applies these hints to prefetchers.\looseness=-1}
    \vspace{-.15in}
    \label{fig:design_overview}
\end{figure}

\subsubsection{Solving feature misleading in prefetcher selection via ML}
Processing these comprehensive features requires a capable algorithm to map them to the IPC-best policy for each phase. 
We adopt a data-driven ML approach. 
ML is highly effective at finding complex, nonlinear relationships between numerous hardware features and IPC. 
By learning this mapping directly, the model evaluates feature importance accurately to make precise prefetcher selections. 
Furthermore, this data-driven approach enables the trained model to generalize effectively, predicting IPC-best policies for unseen workloads.

\subsubsection{Eliminating response lag via phase-level granularity}
Reactive methods incur a response lag by waiting for online feedback to change policies.
GPS eliminates this lag by managing profiling and execution at fixed phases of one million instructions.
Instead of accumulating enough runtime features to make decisions, GPS applies a precomputed policy exactly at the start of each phase.
This immediate application removes online exploration time entirely.
Furthermore, this small phase size ensures timeliness in responding to changing memory access patterns and allows the system to explore a large policy space to reach better performance.
We will evaluate how phase length affects performance in Section~\ref{Sensitivity_Studies}.

\section{GPS Workflow}
Figure~\ref{fig:design_overview} shows the end-to-end deployment workflow of GPS based on a trained ML model. Our introduction to the development of the ML model is in the next Section~\ref{ML_Model}. \textbf{During profiling}, GPS first collects the phase-level feature information, then uses a trained lightweight ML model to map these collected features into phase-level IPC-best policy hints. \textbf{During runtime}, the Policy Hint Walker (PHW) loads each phase's policy hint from DRAM into the policy hint register, which configures the prefetchers to implement the precomputed policy.
This section will introduce these two stages in detail. \looseness=-1

\subsection{Profiling Stage}\label{Profiling}
\begin{figure}[t]
    \centering
    \includegraphics[width=0.8\linewidth]{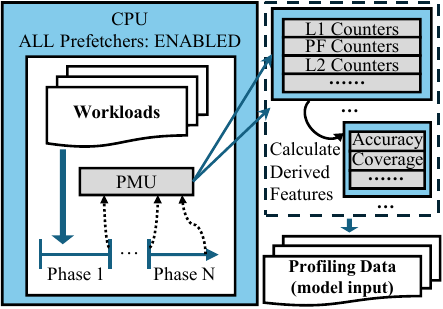}
    \caption{GPS profiling workflow. Program execution is divided into phases with a fixed instruction count. GPS collects microarchitectural features at the end of each phase and computes additional features offline. The original and derived features serve as inputs to the ML model.}
    \vspace{-.1in}
    \label{fig:profiling_process}
\end{figure}
For each target program, GPS performs one profiling run to collect phase-level features and then uses the trained model offline to predict the IPC-best policy for each phase.
A phase is a sequence of one million retired instructions and serves as the basic unit for feature profiling and policy updates.
Figure~\ref{fig:profiling_process} illustrates the process of profiling features.
At the end of each phase, GPS records microarchitectural hardware counters that capture the phase's memory system behavior, prefetcher states, and control flow information. These statistics form the input features to the trained model.
In order to detect the end of each phase, the Performance Monitoring Unit (PMU) tracks the number of retired instructions to identify the current execution phase. When the instruction count exceeds the phase length, a counter overflow interrupt triggers the PMU to record features of each phase. \looseness=-1

\begin{table}[t]
\centering
\caption{Summary of original and derived profiling features.}
\vspace{-.08in}
\label{tab:profiling_features}

\begingroup
\footnotesize
\setlength{\tabcolsep}{2.5pt}
\renewcommand{\arraystretch}{1.10}

\begin{tabular}{|c|@{}c@{}|}
\hline
\textbf{Feature type} &
\begin{tabular}[c]{p{0.57\columnwidth}|c}
{\centering\textbf{Feature details}\par} &
\makebox[0.09\columnwidth][c]{\textbf{Count}}
\end{tabular}
\\
\hline\hline

\textbf{Original} &
\begin{tabular}[c]{p{0.57\columnwidth}|c}
\textbf{Cache and MSHR:}
L1-D, L2, and L3 demand hits, misses, accesses, total miss
latency, MSHR hits, MSHR misses, and total MSHR-miss latency.
&
\makebox[0.09\columnwidth][c]{$21$}
\\
\hline

\textbf{Prefetcher:}
Issued and useful prefetch counts for the four prefetchers.
&
\makebox[0.09\columnwidth][c]{$8$}
\\
\hline

\textbf{Memory bandwidth:}
Read and write bandwidth.
&
\makebox[0.09\columnwidth][c]{$2$}
\\
\hline

\textbf{Branch:}
Branch predictions and mispredictions.
&
\makebox[0.09\columnwidth][c]{$2$}
\\
\hline

\textbf{Subtotal}
&
\makebox[0.09\columnwidth][c]{\textbf{$33$}}
\end{tabular}
\\
\hline\hline

\textbf{Derived} &
\begin{tabular}[c]{p{0.57\columnwidth}|c}
\textbf{Cache and MSHR:}
Demand miss rate, average demand-miss latency, MSHR miss rate,
and average MSHR-miss latency for L1-D, L2, and L3.
&
\makebox[0.09\columnwidth][c]{$12$}
\\
\hline

\textbf{Prefetcher:}
Accuracy and coverage for the four prefetchers.
&
\makebox[0.09\columnwidth][c]{$8$}
\\
\hline

\textbf{Memory bandwidth:}
Total bandwidth utilization.
&
\makebox[0.09\columnwidth][c]{$1$}
\\
\hline

\textbf{Branch:}
Branch misprediction ratio.
&
\makebox[0.09\columnwidth][c]{$1$}
\\
\hline

\textbf{Subtotal}
&
\makebox[0.09\columnwidth][c]{\textbf{$22$}}
\end{tabular}
\\
\hline\hline

\textbf{Total} &
\begin{tabular}[c]{p{0.57\columnwidth}|c}
\textbf{All profiling features}
&
\makebox[0.09\columnwidth][c]{\textbf{$55$}}
\end{tabular}
\\
\hline
\end{tabular}

\endgroup
\vspace{-.2in}
\end{table}

During profiling, GPS enables all prefetchers and sets each to a fixed degree. Although this all-enabled configuration differs from the policy hint configuration used at runtime, it allows each prefetcher to observe the complete demand access stream under the same phase conditions.
The resulting per-prefetcher features provide a more complete view of their behavior and help the ML model map phase behavior to the IPC-best policy.
This all-enabled profiling state is supported on Intel platforms with prefetcher control bits. GPS can collect features under this fixed state~\cite{intel2023sdm,intel_atom_prefetch}.

The features can be categorized into two main types. (1) Original features, which are collected directly from the hardware counters in the PMU. (2) Derived features, which are collected by normalizing some of the original features. Table~\ref{tab:profiling_features} summarizes these features in detail.
Original features can be collected from four aspects of hardware counters. Cache and MSHR counters record hit and miss counts and access latency. Memory bandwidth counters measure off-chip demand traffic. Prefetcher counters record the activity and usefulness of each prefetcher. Branch counters capture branch prediction behavior.
Following a profiling methodology similar to Prophet~\cite{li2025profile}, GPS configures Intel's Processor Event-Based Sampling (PEBS)~\cite{intel2023sdm} to collect original features at the phase level. 
Derived features are computed offline from the original features. They include prefetch accuracy, coverage, cache miss rate, and average miss latency. These normalized features make the same feature easier to compare across phases. In total, GPS collects 33 original features and computes 22 derived features, producing 55 features in total.

\revv{About 20 of the 33 original features can be measured using existing PMU and PEBS events~\cite{intel2023sdm}. These include cache statistics, memory bandwidth, branch statistics, and some MSHR statistics. Remaining features such as prefetcher counters can be collected by adding simple counters, tags, and timestamps to existing hardware structures, with low additional hardware cost. Similar PMU/PEBS-based profiling has been shown to have low overhead~\cite{li2025profile}.}

After profiling, GPS orders the features for each phase and applies the trained model offline. The model predicts one policy for each phase and produces an ordered policy hint sequence.
The policy sequence is stored in memory for use in the running stage. No feature derivation or model inference occurs during online execution.

\subsection{Running Stage}\label{sec::Running_Stage}
The running stage executes the target program using precomputed policy hints without online model inference. Before execution, the operating system loads the hint table into DRAM and initializes the PHW with its base address. Each hint specifies up to two active prefetchers and their corresponding degrees.\looseness=-1

During execution, the PMU counts retired instructions, and the PHW monitors this count directly. PHW uses a 512-bit on-chip buffer to cache 32 consecutive hints and a 5-bit index to select them sequentially. The next hint is prepared in advance, allowing the PHW to update the active policy register directly at each phase boundary. The hint determines which prefetchers receive demand requests, which prefetchers may issue prefetch requests, and the degree of each active prefetcher.

Overall, GPS requires only one on-chip hint lookup per phase and one DRAM fetch every 32 phases. It avoids online feature collection, model inference, and policy exploration, enabling timely policy updates with minimal runtime overhead. Section~\ref{sec:On-chip_Overhead} details the overhead.

\revv{GPS can generalize different inputs for the same workload in two levels. First, GPS can generate and apply new hints for unseen inputs without retraining the model. Second, GPS can directly reuse policy hints generated for one input when running the same program with another input.}
\revv{Direct hint reuse generally assumes that memory access patterns and control flow do not change substantially across inputs. More specifically, reused hints can remain effective when the two inputs have similar average profiling features. We evaluate both scenarios in Section~\ref{Different_Program_Inputs}.}
\section{GPS ML Model}\label{ML_Model}
This section describes how we construct and train the policy prediction ML model used to generate IPC-best policy hints. It covers the construction of the policy space, the design of the ML model, and the training procedure.
Section~\ref{sec:overthead_studies} discusses the cost of ML model training and inference. 

\subsection{Policy Space}\label{Policy_Space}
We construct a policy space and derive IPC-related labels to build the training dataset.
The policy space contains all single prefetcher policies, each with one active prefetcher, and a sampled subset of double prefetcher policies, each with two active prefetchers.
We sample double prefetcher policies because exhausting all double prefetcher policy combinations significantly increases the policy space while providing negligible additional speedup (as evaluated in Section~\ref{section:candpolicy_space}).
Each policy in the policy space is executed on the target platform, and its phase-level IPC values are recorded to construct the training labels.\looseness=-1

\begin{figure}[t]
    \centering
    \includegraphics[width=0.95\linewidth]{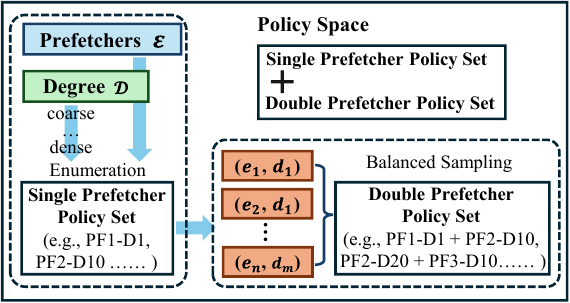}
    \vspace{-.05in}
    \caption{Policy space construction. Single prefetcher policies are enumerated, and double prefetcher policies are sampled from feasible pairs of single prefetcher policies.}
    \vspace{-.15in}
    \label{fig:policy_space}
\end{figure}

\textbf{(1) Policy Space Construction}: Figure~\ref{fig:policy_space} shows how to construct the policy space in two levels. It first enumerates all single prefetcher policies. Let $\mathcal{E}$ be the set of candidate prefetchers and $\mathcal{D}$ be the degree values supported by prefetcher $e$. A single prefetcher policy is defined as $(e,d)$, where $e\in\mathcal{E}$ and $d\in\mathcal{D}$.\looseness=-1

The policy space includes both single and double prefetcher policies because a phase may contain multiple demand access patterns. Modern multi-prefetcher systems combine specialized prefetchers that target different patterns.
A single prefetcher may not capture all memory access patterns in a complex phase, whereas two complementary prefetchers can provide broader coverage.

A policy specifies demand request allocation, prefetch request filtering, and prefetch degree. In a single prefetcher policy, all demand accesses are sent to one prefetcher, and only that prefetcher may issue requests. In a double prefetcher policy, both active prefetchers receive the demand stream and issue requests at their configured degrees.

Enumerating all feasible double prefetcher policies would cause the policy space to grow rapidly. We therefore construct a bounded set using coverage-balanced pair selection. A feasible double prefetcher policy is written as $((e_1,d_1),(e_2,d_2))$, where $e_1\neq e_2$. Let $\mathcal{F}$ denote all feasible pairs and $\alpha$ denote the sampling rate.
We select $\alpha|\mathcal{F}|$ pairs while balancing how often each single prefetcher policy appears. The final policy space contains all single prefetcher policies and the sampled double prefetcher policies.

\textbf{(2) Labeling}:
Training labels are derived from the IPC measured for every policy of each phase. We execute every policy on the target platform and record its IPC of each phase.
We construct training labels in two formats: a continuous policy score for the regression task and a binary top 10\% policy label for the classification task.

To construct the continuous policy score, we normalize each policy's IPC as: Normalized IPC $=$ Policy's IPC  $/$ Highest IPC in this phase. We refer to this \emph{normalized IPC} as the policy score. A higher policy score indicates better performance within the same phase, and the best policy receives a score of one. This normalization removes variations in absolute IPC scales across phases while preserving the relative performance and ranking of policies within each phase.

To construct the binary top 10\% policy label, we rank all candidate policies by their IPC separately within each phase. Policies ranked in the top 10\% are assigned a positive label, whereas all remaining policies are assigned a negative label.

As for features, we collect features of each phase using the profiling method described in Section~\ref{Profiling}. Features are paired with the policy scores and binary labels of all policies in the corresponding phase to form the training dataset.\looseness=-1

\begin{figure}[t]
    \centering
    \includegraphics[width=0.9\linewidth]{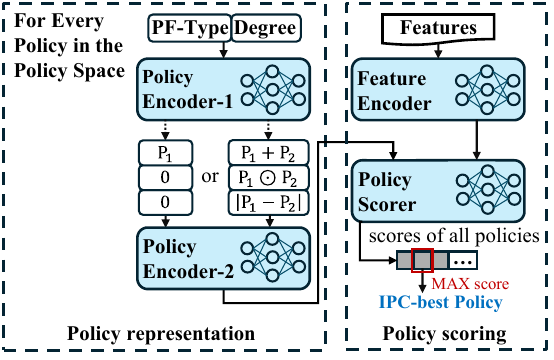}
    \vspace{-.1in}
    \caption{ML model structure. Policy representation encodes each policy's prefetcher types and degrees. Policy scoring combines this representation with phase features and selects the policy with the highest score. End-to-end training jointly updates all components.}
    \vspace{-.15in}
    \label{fig:model_construction}
\end{figure}

\subsection{ML Model Construction}\label{Model_Construction}
Figure~\ref{fig:model_construction} shows the GPS ML model structure.
For the given features of a phase, the model evaluates the phase features with each policy and outputs a score separately, ranks all predicted scores, and selects the policy with the highest score as its output IPC-best policy.

Instead of directly mapping phase features to a policy through conventional multiclass classification, our model encodes policy and phase information separately. It then combines their representations to score each policy.
Accordingly, the model consists of two parts: policy representation and policy scoring. As shown in Figure~\ref{fig:model_construction}, these parts are implemented by four lightweight MLPs. These MLPs are named according to their functions: Policy Encoder-1, Policy Encoder-2, Feature Encoder, and Policy Scorer.

\textbf{Policy representation.}
First, Policy Encoder-1 encodes the type and degree of a single policy. The result of Policy Encoder-1 is denoted by \(P\). This can keep the different meanings of prefetcher type and degree.
Second, Policy Encoder-2 calculates the policy representation.
For a single policy, it uses \(P_1\) as input.
For a double policy, it takes the results of a series of symmetric operations on \(P_1\) and \(P_2\) as input.
This design preserves the structural relationship between single and double policies.
The Policy Encoder-1 output \(P\) for a single policy is reused whenever that policy appears in a double policy. This reuse allows related policies to share learned information.

\textbf{Policy scoring.}
The Feature Encoder encodes the input features of each phase into a feature representation. The Policy Scorer combines this feature representation with each policy representation to predict a score for every policy.
Each score estimates how well a policy matches the current phase because it incorporates information from both the phase and the policy. The model then selects the policy with the highest score as the predicted policy.

\subsection{ML Model Training}\label{Model_Training}
We use end-to-end training for the entire model in two sequential stages, jointly updating all four MLPs in both stages.
Stage~1 formulates policy score prediction as a regression task, whereas Stage~2 formulates top 10\% policy identification as a binary classification task. The normalized IPC captures detailed performance differences, while the top $10\%$ label directs training toward better policies.

\textbf{Stage 1: Policy score regression.}
We use the normalized IPC as the regression target and train the model to predict its policy score. We use the \(L_2\) loss~\cite{goodfellow2016deep} to penalize the difference between the predicted and measured policy scores. 

\textbf{Stage 2: Top 10\% policy classification.}
Stage~2 is initialized with the parameters learned in Stage~1. For each phase, policies in the top 10\% are labeled as positive, while the remaining policies are labeled as negative. We use the binary cross-entropy loss~\cite{goodfellow2016deep} to train the model to predict whether a policy belongs to the top 10\%.

\begin{figure*}[t!]
    \centering
    \includegraphics[width=\linewidth]{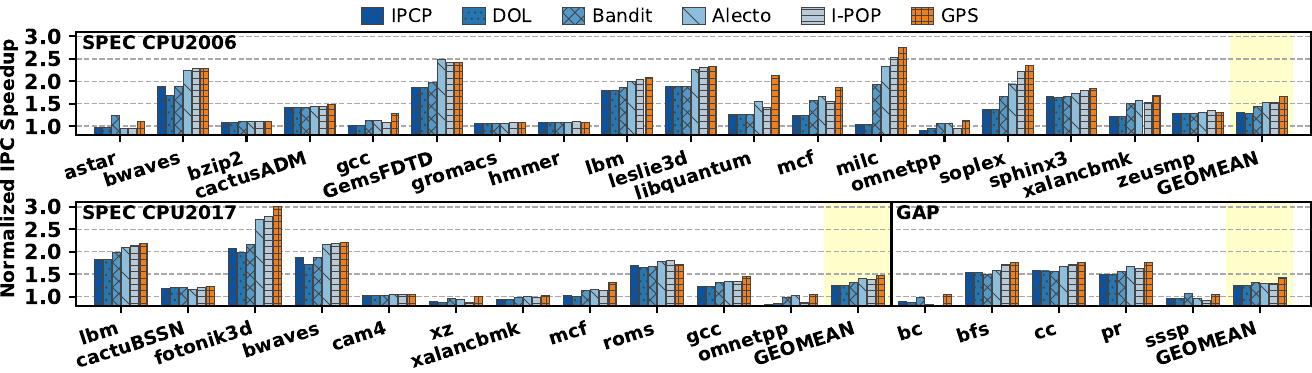}
    \vspace{-.2in}
    \caption{IPC speedup normalized to no prefetching for memory-intensive workloads in SPEC06, SPEC17, and GAP. Their geometric-mean speedups are computed separately and highlighted with a light yellow background. All prefetcher selection methods use the same set of candidate prefetchers: GS~\cite{pakalapati_bouquet_2020} + CS~\cite{pakalapati_bouquet_2020} + CPLX~\cite{pakalapati_bouquet_2020} + Berti~\cite{navarro-torres_berti_2022}.}
    \vspace{-.2in}
    \label{fig:speedup_all}
\end{figure*}
\section{Experimental Methodology}\label{sec:exep_method}

\begin{table}[t!]
\centering
\caption{Simulation System configuration.}
\vspace{-.1in}
\label{tab:sys_confg}
\begin{tabular}{|l|l|}
\hline
\textbf{Module} & \textbf{Configuration} \\
\hline
\hline
\textbf{Core} & fetchWidth: 5, decodeWidth: 5 \\
& issueWidth: 10, commitWidth: 10\\
& IQ Entry: 120, LQ Entry: 85\\
& SQ Entry: 90, ROB Entry: 288\\
\hline
\textbf{\revv{TLB}} & \revv{L1 ITLB/DTLB: 64 entries, 4-way} \\
& \revv{1 cycle lookup latency} \\
& \revv{L2 STLB: 1536 entries, 12-way} \\
& \revv{8 cycles lookup latency} \\
& \revv{Page size: 4KB} \\
\hline
\textbf{L1 I/D Cache} & 64KB, 4-way, 64Byte Line \\
& tag and data latency 4 cycles\\
& PLRU, 16 MSHRs \\
\hline
\textbf{L2 Cache} & 512KB, 8-way, 64Byte Line \\
& tag and data latency 9 cycles\\
& PLRU, mostly\_inclusive, 32 MSHRs \\
\hline
\textbf{L3 cache} & 2MB, 16-way, 64Byte Line \\
& tag and data latency 20 cycles\\
& CHAR~\cite{chaudhuri2012introducing}, mostly\_exclusive, 36 MSHRs \\
\hline
\textbf{DRAM} & LPDDR5\_5500\_1x16\_BG\_BL32 \\
\hline
\end{tabular}
\vspace{-.1in}
\end{table}

\subsection{System Configuration}
We utilize gem5~\cite{binkert2011gem5} to evaluate GPS. PEBS is the intended profiling interface for real deployment, whereas gem5 is our experimental platform for architectural evaluation and dataset collection. The system configuration is listed in Table~\ref{tab:sys_confg}, which is very similar to Triangel~\cite{ainsworth_triangel_2024}. To obtain phase-level runtime features, we modify the gem5 statistics mechanism.
GPS implements the Policy Encoder-1, Policy Encoder-2, Feature Encoder, and Policy Scorer as four jointly trained MLPs~\cite{goodfellow2016deep}. Their layer dimensions are 
(128, 256, 128, 64), (192, 256, 128, 64), (55, 256, 128, 64), and (128, 256, 128, 64, 1), respectively. All hidden layers use
ReLU activation and a dropout rate of 0.2.
GPS trains both stages for 50 epochs with a batch size of 32.
Then, we use the trained ML model to process the features and generate policy hints. 

\subsection{Evaluation Methodology}\label{Evaluation_Methodology}
We evaluate GPS using selected memory-intensive workloads from the SPEC CPU benchmark and graph-processing benchmark.
To assess workload performance, we use the SimPoint technique~\cite{sherwood2002automatically} to select simulated checkpoints for each workload. Each checkpoint contains a warm-up phase involving 100M instructions followed by a simulation of the subsequent 100M instructions.
We obtain the final performance for a benchmark by applying a weighted average to the cycles per instruction (CPI) across all checkpoints.

To evaluate the real-world performance of the GPS prefetcher, we perform workload-level five-fold cross-validation~\cite{kohavi1995study} on all workloads.
We randomly partition the workloads into five folds, using four folds for training and the remaining fold for testing in each cross-validation round.
All SimPoint checkpoints belonging to the same workload are kept together in either the training or testing split.
Thus, every test phase comes from a workload that is entirely unseen during model training.
This ensures GPS's applicability in realistic scenarios, i.e., the actual performance of the ML-derived policy hints on unseen workloads.

We compare GPS with other state-of-the-art composite prefetchers, including IPCP~\cite{pakalapati_bouquet_2020}, DOL~\cite{kondguli_division_2018}, Bandit~\cite{gerogiannis_micro-armed_2023}, Alecto~\cite{li2025integrating}, and I-POP~\cite{lin2026pop}.
These prefetchers are configured as L1-D cache prefetchers, selecting prefetching policies for four different hardware prefetchers: CS~\cite{pakalapati_bouquet_2020}, GS~\cite{pakalapati_bouquet_2020}, CPLX~\cite{pakalapati_bouquet_2020}, and Berti~\cite{navarro-torres_berti_2022}.
CS targets per-instruction constant strides, GS targets dense global streams, and CPLX targets repeated but non-constant stride patterns. Berti is a local-delta prefetcher that selects timely, high-coverage deltas for each instruction. Together, these prefetchers cover regular streams, constant strides, complex strides, and local delta patterns.
\revv{All four prefetchers allow cross-page prefetching. Since our system has no dedicated TLB prefetcher, we use this capability to also prefetch address translations~\cite{vavouliotis2025cross, vavouliotis2022page, liu2024planaria}.}
The L2 cache prefetcher is configured as Triage~\cite{wu_triage_2019}.

\section{Evaluation}
\subsection{Single-Core Performance}\label{Single-Core_Performance}
Figure~\ref{fig:speedup_all} compares the IPC speedup of the GPS prefetcher against state-of-the-art prefetcher selection methods, including IPCP, DOL, Bandit, Alecto, and I-POP.
The results are normalized to a no-prefetching baseline across the SPEC CPU2006~\cite{henning2006spec}, SPEC CPU2017~\cite{spec2017}, and GAP~\cite{beamer2015gap} benchmark suites.
On average, for SPEC06 benchmarks, GPS outperforms I-POP by 12.14\%, Alecto by 11.85\%, Bandit by 21.39\%, DOL by 36.17\%, and IPCP by 35.37\%.
For SPEC17 benchmarks, GPS outperforms I-POP by 8.34\%, Alecto by 7.01\%, Bandit by 13.94\%, DOL by 22.08\%, and IPCP by 20.84\%. 
For GAP benchmarks, GPS outperforms I-POP by 14.17\%, Alecto by 14.06\%, Bandit by 12.03\%, DOL by 17.78\%, and IPCP by 17.22\%.
Across the three benchmark suites, GPS achieves a 7.01\%--14.17\% performance improvement even when compared against the most recent state-of-the-art prefetcher selection algorithms, Alecto and I-POP.
The performance advantage of GPS comes from using profiling and an ML model.
GPS avoids online exploration and response lag by applying profiled policy hints at phase boundaries.

\begin{figure}
    \centering
    \includegraphics[width=0.8\linewidth]{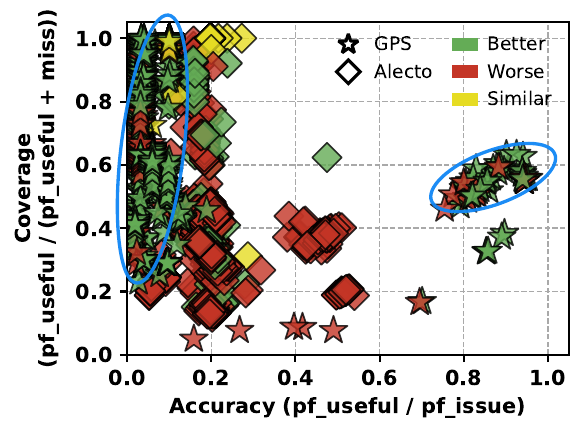}
    \vspace{-.1in}
    \caption{Phase-level prefetch accuracy and coverage for \textit{milc}. Each point represents the performance of a prefetcher in one phase. Green, red, and yellow indicate higher, lower, and within 1\% IPC compared to the alternative, respectively. Blue ellipses highlight the two main GPS clusters.\looseness=-1}
    \vspace{-.25in}
    \label{fig:milc_acc_cov}
\end{figure}

Figure~\ref{fig:milc_acc_cov} compares the phase behavior distributions of GPS and Alecto on \textit{milc}.
While Alecto, a mapping algorithm primarily driven by accuracy, mainly clusters in the 0.2 to 0.5 accuracy range, GPS forms two clusters at 0 to 0.2 and 0.8 to 1.0.
Despite lower accuracy, GPS achieves higher IPC in the first cluster, as shown by the dominant green points.
GPS also identifies a high accuracy region (0.8 to 1.0) missed by Alecto.
These results show that accuracy alone does not always reflect IPC, while GPS avoids misleading features and finds better policies by exploring a larger policy space offline.

\begin{figure}
    \centering
    \includegraphics[width=1\linewidth]{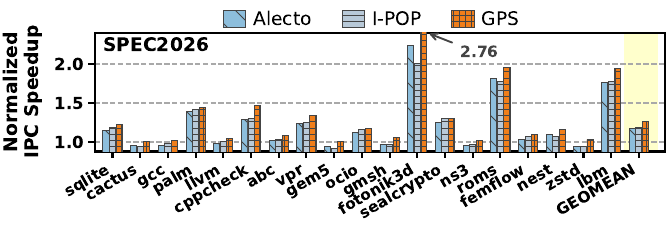}
    \vspace{-.25in}
    \caption{\revv{IPC speedup normalized to no prefetching for memory-intensive workloads in SPEC2026.}}
    \vspace{-.15in}
    \label{fig:speedup_spec2026}
\end{figure}

\revv{Figure~\ref{fig:speedup_spec2026} further evaluates GPS on 19 memory-intensive SPEC CPU 2026 workloads~\cite{madhav2026spec}.
GPS improves geometric-mean IPC by 7.9\% over Alecto and 7.3\% over I-POP.
Across individual workloads, the improvements range from 4.0--23.2\% over Alecto and 0.2--37.2\% over I-POP.
These results demonstrate the robustness of GPS across diverse memory-intensive workloads in SPEC CPU 2026.}\looseness=-1

\begin{figure}
    \centering
    \includegraphics[width=1\linewidth]{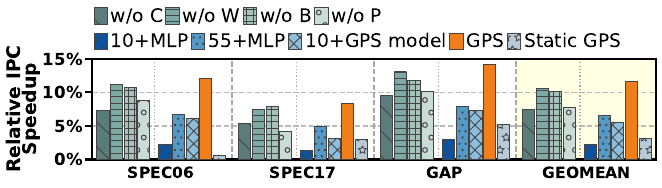}
    \vspace{-.2in}
    \caption{\revv{Relative IPC speedup of GPS performance breakdown and feature ablation over I-POP. C, W, B, and P denote cache and MSHR, bandwidth, branch, and prefetcher features. 10 and 55 denote I-POP and GPS feature sets. Static GPS is a static profile-guided baseline using one fixed policy per workload.}}
    \vspace{-.2in}
    \label{fig:feature_ablation}
\end{figure}

\subsection{\revv{GPS Performance Breakdown}}\label{perf_breakdown}
\revv{Figure~\ref{fig:feature_ablation} decomposes the performance gain of GPS. Static GPS is a static profile-guided baseline that uses a single fixed policy for each workload, without phase-level switching. \textit{10+MLP} and \textit{55+MLP} use I-POP's 10 features and GPS's full 55 features, respectively, with a simple MLP; the \textit{10+GPS model} uses the 10 features with GPS's structured policy model (discussed in Section~\ref{ML_Model}), while GPS uses the original configuration.
We further conduct a feature ablation study by removing each feature group from full GPS following the categories in Table~\ref{tab:profiling_features}. C, W, B, and P denote cache and MSHR, memory bandwidth, branch, and prefetcher features, respectively; for each group, both the original and corresponding derived features are removed.}\looseness=-1

\begin{figure}[t]
    \centering
    \includegraphics[width=1\linewidth]{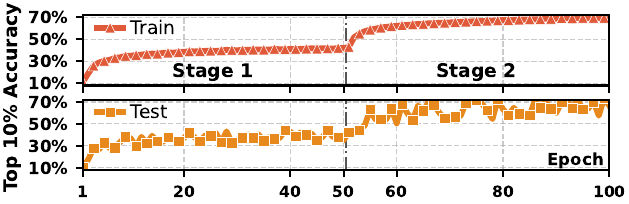}
    \vspace{-.2in}
    \caption{Top $10\%$ training and testing accuracy across the two training stages in five-fold cross-validation.}
    \vspace{-.2in}
    \label{fig:modelacc}
\end{figure}

\revv{Overall, the ablation results show that GPS performance gain comes from multiple components rather than a single dominant factor.
\textbf{First}, using the richer 55 feature set increases the gain from 2.22\% to 6.50\% under the simple MLP, while replacing the simple MLP with the structured GPS ML model increases it from 2.22\% to 5.55\% with the same 10 features. Combining both achieves the largest gain, indicating that richer features and structured policy mapping make complementary contributions.
\textbf{Second}, the feature ablation shows that cache and MSHR, along with prefetcher features, provide the largest marginal benefits: removing them reduces the gain to 7.44\% and 7.75\%, respectively, whereas bandwidth and branch features have smaller effects.
\textbf{Finally}, Static GPS achieves only 3.10\%, showing that selecting a static policy per workload is insufficient and that phase-level policy selection is an important source of GPS advantage.}
\vspace{-.1in}

\subsection{ML Model Evaluation}\label{section:model_eval}

\subsubsection{Prediction Accuracy}
Figure~\ref{fig:modelacc} reports the top $10\%$ training and testing accuracy across the two training stages. A prediction is correct when the selected policy belongs to the top $10\%$ candidate policies ranked by measured IPC.
The first 50 epochs correspond to policy score regression in Stage 1, while the next 50 epochs correspond to binary classification of top-10\% policies in Stage 2.

The learning curves demonstrate smooth convergence without severe overfitting. The model ultimately reaches a top 10\% test accuracy of nearly 66\%.

Top $10\%$ accuracy is meaningful because the highest-ranked policies often provide similar IPC gains. We measure the performance of the top $N\%$ policies using the IPC speedup of the lowest policy within that group. According to our evaluation across SPEC06, SPEC17, and GAP, compared with the top $1\%$, the IPC speedup drops by only 1.9\% for the top $10\%$, versus 5.1\% and 10.8\% for the top $20\%$ and $40\%$, respectively. Thus, the top $10\%$ threshold allows multiple valid predictions while excluding policies with substantial IPC loss.

\subsubsection{Feature Importance}
We use SHAP values~\cite{lundberg2017unified} to quantify the importance of input features in the GPS
model's policy decisions. As shown in Table~\ref{tab:shap_features}, the historical coverage and accuracy of candidate prefetchers (e.g., GS and Berti), together with cache latency and memory read bandwidth, receive the highest SHAP scores. This indicates that the model relies on prefetcher effectiveness and memory hierarchy state when selecting policies.\looseness=-1

\begin{table}[t]
    \centering
    \caption{Top 10 most important features ranked by SHAP values, evaluated using five-fold cross-validation.}
    \label{tab:shap_features}
    \begin{tabular}{|l|c|c|}
        \hline
        \textbf{Feature} & \textbf{Mean SHAP} & \textbf{Std. Dev.} \\
        \hline
        \hline
            gs pf.accuracy	    & 0.1190	& 0.0165       \\
            berti pf.coverage	& 0.0946	& 0.0313       \\
            gs pf.coverage	    & 0.0877	& 0.0218       \\
            L2 MissRate	        & 0.0826	& 0.0082       \\
            L2 MSHRMissRate	    & 0.0767	& 0.0189       \\
            cplx pf.accuracy	& 0.0745	& 0.0132       \\
            L2 AvgMissLat	    & 0.0738	& 0.0143       \\
            Mem Ctrl Rd BW	    & 0.0723	& 0.0147       \\
            L3 AvgMSHRMissLat	& 0.0723	& 0.0109       \\
            L3 MissRate	        & 0.0714	& 0.0041       \\
        \hline
    \end{tabular}
\vspace{-.15in}
\end{table}

\subsection{Generality}
\subsubsection{Prefetcher Set Generality}\label{Prefetcher_Set_Generality}
\begin{figure}[t]
    \centering
    \includegraphics[width=1\linewidth]{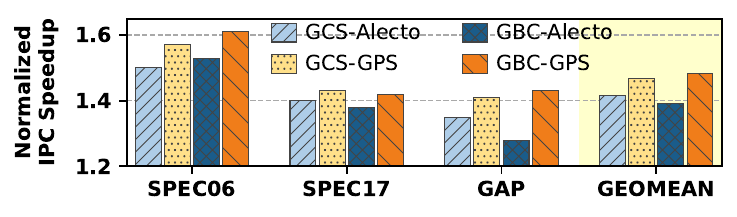}
    \vspace{-.2in}
    \caption{IPC speedup when scheduling different combinations of prefetchers, normalized to the no-prefetching baseline. We evaluate two kinds of combinations: \textbf{GCS} (GS~\cite{pakalapati_bouquet_2020} + CS~\cite{pakalapati_bouquet_2020} + SMS~\cite{somogyi_spatial_2006}) and \textbf{GBC} (GS~\cite{pakalapati_bouquet_2020} + Berti~\cite{navarro-torres_berti_2022} + CPLX~\cite{pakalapati_bouquet_2020}).}
    \vspace{-.2in}
    \label{fig:speedup_diversepf}
\end{figure}

Figure~\ref{fig:speedup_diversepf} evaluates the normalized IPC speedups when GPS and Alecto manage two different combinations of prefetchers.

On average, GPS outperforms Alecto by 5.30\% under the GCS configuration and significantly widens this lead to 9.12\% under the GBC configuration.

This gap highlights a key difference between the two kinds of prefetching selection algorithms. 
Aggressive prefetchers such as CPLX can handle complex access patterns but may also increase bandwidth pressure and cache interference. Because Alecto relies on local, reactive signals derived from recent prefetch hits, it cannot fully account for these system-level effects.\looseness=-1

Instead of making local reactive decisions, GPS chooses the prefetching policy in advance while keeping bandwidth and cache pollution under control. This shows GPS's architectural advantage: its predictive model can outperform fixed hardware trackers and better unlock the performance potential of complex prefetcher combinations.

\subsubsection{\revv{Different Workload Inputs}}\label{Different_Program_Inputs}

\begin{figure}[t]
    \centering
    \includegraphics[width=1\linewidth]{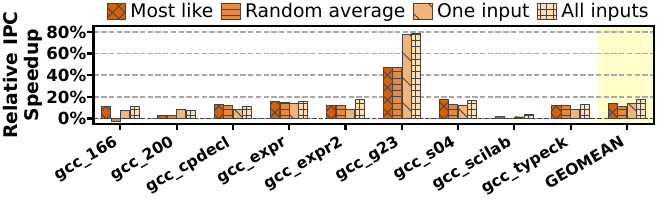}
    \vspace{-.2in}
    \caption{\revv{Relative IPC speedup over I-POP across nine \textit{gcc} inputs, evaluating GPS generalization to unseen inputs.}}
    \label{fig:gccinput_gen}
    \vspace{-.2in}
\end{figure}

\revv{Figure~\ref{fig:gccinput_gen} evaluates GPS's input generalization at two levels: generating new hints without model retraining and directly reusing existing hints.}
\revv{For the first level, \emph{All inputs} is an ``ideal'' reference trained on all nine \textit{gcc} inputs. For each target input, \emph{One input} averages the performance of eight models, each trained on one of the other inputs. These models generate new hints from the target's profiled features without retraining.}
\revv{For direct hint reuse, \emph{Most like} reuses hints from the other input with the most similar average profiling features (minimum Euclidean distance). \emph{Random average} reports the average performance of reusing hints randomly selected from other inputs, representing reuse without similarity-based selection. The geometric-mean IPC gains are 14.0\% for \emph{One input} and 17.3\% for \emph{All inputs}. Direct hint reuse also achieves gains of 13.7\% for \emph{Most like} and 11.3\% for \emph{Random average}. These results show that GPS generalizes to unseen inputs without retraining and enables effective hint reuse across inputs, especially those with similar behavior.}\looseness=-1

\subsubsection{\revv{Multicore Performance}}\label{Multicore_Perf}
\begin{figure}[t]
    \centering
    \includegraphics[width=1\linewidth]{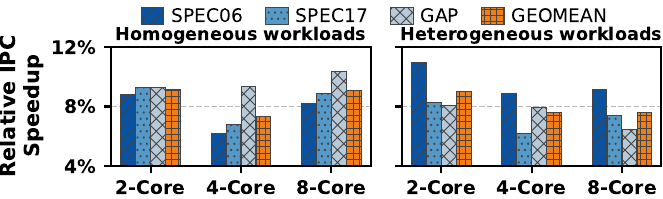}
    \vspace{-.2in}
    \caption{\revv{Relative IPC speedup of GPS over I-POP in homogeneous and heterogeneous multicore systems.}}
    \vspace{-.15in}
    \label{fig:speedup_multicore}
\end{figure}

\revv{Figure~\ref{fig:speedup_multicore} compares GPS with I-POP on homogeneous and heterogeneous multicore workloads.
For homogeneous workloads, GPS directly reuses the generated hints and improves IPC by 9.1\%, 7.4\%, and 9.1\% with 2, 4, and 8 cores, respectively.
For heterogeneous workloads, GPS still outperforms I-POP by 9.0\%, 7.6\%, and 7.6\%.
These results show that GPS maintains its advantage across different workload mixes and core counts.}
\vspace{-.1in}

\subsection{Sensitivity Studies}\label{Sensitivity_Studies}
\subsubsection{Phase Length}
\begin{figure}[t]
    \centering
    \includegraphics[width=1\linewidth]{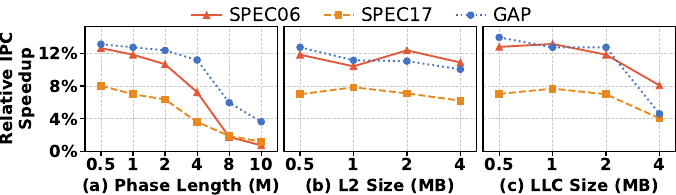}
    \vspace{-.2in}
    \caption{Relative IPC speedup of GPS over I-POP with varying (a) phase length, \revv{(b) L2 cache capacity,} (c) LLC capacity}
    \vspace{-.15in}
    \label{fig:sen}
\end{figure}

Figure~\ref{fig:sen}(a) evaluates the sensitivity of GPS's relative IPC gains over I-POP to phase length. At 0.5M instructions, GPS achieves the largest gains.
Longer phases merge distinct behaviors and reduce policy update opportunities, so GPS gradually loses its advantage at 8M and beyond. GPS therefore adopts a phase length of 1M instructions as a practical trade-off: its gains remain within 1\% of the 0.5M results across all suites, while requiring only half as many policy updates.

\begin{figure}[t]
    \centering
    \includegraphics[width=1\linewidth]{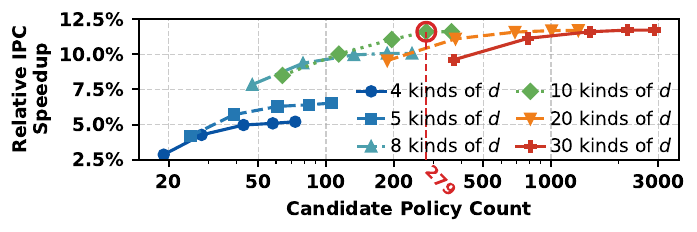}
    \vspace{-.25in}
    \caption{Policy space exploration. The curves show relative IPC speedup as the policy space size increases under different numbers of chosen degrees and sampling ratios. The red circle marks the selected setting.}
    \vspace{-.15in}
    \label{fig:policy_space_budget}
\end{figure}

\subsubsection{\revv{Cache Capacity}}\label{Sensitivity_Studies_Cache}
\revv{Figure~\ref{fig:sen}(b) and Figure~\ref{fig:sen}(c) evaluate GPS across L2 and LLC sizes from 512KB to 4MB.
Across L2 sizes, GPS maintains relatively stable positive speedups, with only a slight reduction at 4MB.
For LLC, the performance gain decreases more noticeably as capacity increases, since larger LLCs reduce cache misses and thus leave fewer memory stalls for prefetching to hide.
However, GPS maintains positive speedups across all evaluated cache configurations, demonstrating its effectiveness across different cache capacities.}

\subsubsection{Policy Space Size}\label{section:candpolicy_space}
Figure~\ref{fig:policy_space_budget} explores the different policy space sizes generated by varying $\mathcal{D}$ and the double policy sampling ratio $\alpha$. For each setting, we report the geometric-mean IPC speedup of GPS over I-POP.
For each evaluated degree grid $\mathcal{D}$, CS, GS, and CPLX support every degree $d \in \mathcal{D}$. Berti has a fixed degree of $d=1$ and is not subject to degree control. 
Relative IPC speedup increases with the size of the policy space but gradually saturates. GPS therefore selects the red-circled setting, where $\mathcal{D}=\{1,2,3,4,5,10,15,20,25,30\}$ and $\alpha=0.75$. This setting contains 279 policies in the policy space and retains most of the observed IPC benefit with a manageable offline evaluation cost.
\vspace{-.1in}

\subsection{Overhead Studies}\label{sec:overthead_studies}

\begin{table}[t]
    \centering
    \caption{\revv{GPS on-chip storage overhead.}}
    \vspace{-.08in}
    \label{tab:phw_overhead}
    \begin{tabular}{|l|c|c|}
        \hline
        \textbf{Component} & \textbf{Count $\times$ Width} & \textbf{Storage} \\
        \hline
        \hline
        Hint buffer                & $32 \times 16$ bits          & 512 bits \\
        Address and phase tracking & $3 \times 64 + 32 + 5$ bits & 229 bits \\
        Policy registers           & $2 \times 16$ bits           & 32 bits \\
        Control flags              & $5 \times 1$ bit             & 5 bits \\
        \hline
        \textbf{Total}             & -- & \textbf{778 bits} \\
        \hline
    \end{tabular}
    \vspace{-.15in}
\end{table}

\subsubsection{\revv{On-chip and Off-chip Overhead}}\label{sec:On-chip_Overhead}
\revv{Table~\ref{tab:phw_overhead} summarizes the on-chip storage overhead required to support GPS.
The 512-bit hint buffer caches 32 consecutive 16-bit policy hints, reducing memory accesses.
The 229-bit address and phase tracking registers include two 64-bit hint table pointers, a 64-bit hint table boundary, a 32-bit phase length, and a 5-bit buffer index that selects one of the 32 cached policy hints. They locate the next hint and determine when to update policies using the existing PMU instruction counter.
Two 16-bit policy registers hold the active and upcoming policies, enabling immediate updates at phase boundaries.
Five control bits manage hint validity and memory fetch status.
In total, GPS requires only 778 bits of on-chip storage, equivalent to 0.15\% of the 64-KB L1-D cache capacity, plus simple control logic. Thus, the additional on-chip hardware overhead is minimal.
For off-chip storage, each hint contains two 3-bit prefetcher IDs and two 5-bit degree settings, totaling 16 bits.
Since hints are accessed sequentially, no phase ID is needed. With one hint per million instructions, GPS requires only 2 KB per billion instructions.}

\subsubsection{Energy Overhead}
\begin{figure}[t]
    \centering
    \includegraphics[width=1\linewidth]{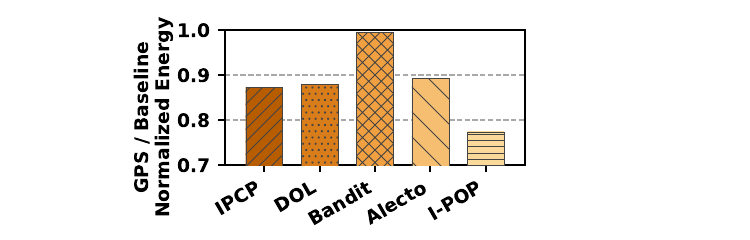}
    \vspace{-.25in}
    \caption{Normalized energy consumption of the memory hierarchy for GPS relative to five baselines.}
    \vspace{-.2in}
    \label{fig:energy_relative}
\end{figure}

We use CACTI~\cite{li2011cacti} at a 22nm node to evaluate the memory hierarchy energy, setting the DRAM access cost to 25$\times$ the LLC access cost, as in Triangel~\cite{ainsworth_triangel_2024}. Our results are shown in Figure~\ref{fig:energy_relative}. Overall, GPS reduces total energy by 12.7\% compared to IPCP, 12.0\% compared to DOL, 0.5\% compared to Bandit, 10.8\% compared to Alecto, and 22.7\% compared to I-POP.

\subsubsection{Profiling Overhead}~\label{sec::profiling_overhead}
GPS collects 33 original features per phase, most of which are simple event counts with low collection overhead.
About 20 features are supported by commodity PMU/PEBS mechanisms~\cite{intel2023sdm}, while the rest require simple counters, as discussed in Section~\ref{Profiling}.
Cache hierarchy latency measurements require three PEBS events.
Prior work~\cite{li2025profile} reports less than 5\% performance overhead when sampling four PEBS events, suggesting similarly low sampling overhead for GPS.
\revv{
We assume that one profiling run takes 5\% longer than an I-POP run. GPS improves IPC by about 12\% over I-POP, reducing execution time by about 11\% per run. Thus, about 10 GPS runs are needed to cover the profiling cost.\looseness=-1}

\subsubsection{ML Model Training and Inference Cost}
GPS performs label generation, training, and inference entirely offline, adding no runtime ML overhead.
Evaluating 279 policies over a benchmark suite of 100 trillion instructions takes about 18.5 hours per policy on a 3-GHz CPU at 0.5 IPC, or 1.7 days with 128 parallel workers.
GPS trains four small MLPs for 50 epochs per stage.
During inference, policy representations can be reused across all phases, keeping inference cost low.\looseness=-1

\section{\revv{Related Works}}\label{sec::Related_Works}
\begin{figure}[t]
    \centering
    \includegraphics[width=1\linewidth]{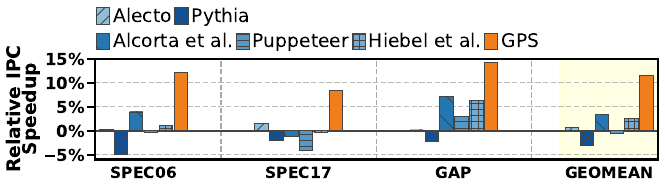}
    \vspace{-.2in}
    \caption{\revv{Relative IPC speedup of different works over I-POP. Alcorta et al.~\cite{alcorta2024characterizing}, Puppeteer~\cite{eris2022puppeteer}, and Hiebel et al.~\cite{hiebel2019machine} use their respective feature sets with the GPS model.}}
    \vspace{-.2in}
    \label{fig:prefetcher_comparison_relative}
\end{figure}

\revv{In this section, we discuss prior work related to GPS and highlight the key differences.}
\subsection{\revv{Profile-Guided Prefetching}}

\subsubsection{\revv{Profile-Guided Prefetcher Selection}}

\revv{Works like Liao et al.~\cite{liao2009machine}, Rahman et al.~\cite{rahman2015maximizing}, Hiebel et al.~\cite{hiebel2019machine}, Puppeteer~\cite{eris2022puppeteer}, and Alcorta et al.~\cite{alcorta2024characterizing}
are most closely related to GPS. However, their limitations are \textbf{(1) applying uniform policies for all execution phases}~\cite{liao2009machine, rahman2015maximizing}: they cannot adapt to pattern changes during program execution; \textbf{(2) using incomplete input features}~\cite{hiebel2019machine, eris2022puppeteer, alcorta2024characterizing}: Their limited feature sets can also restrict selection performance.
To evaluate this limitation, Figure~\ref{fig:prefetcher_comparison_relative} compares the 11, 6, and 11 features used by Alcorta et al., Puppeteer, and Hiebel et al., respectively, using the same GPS model.
The corresponding legend entries denote GPS variants using these feature sets.
Their geometric-mean normalized IPC gains are 8.26\%, 12.07\%, and 9.10\% lower than GPS, respectively.
These results highlight the importance of richer and appropriate features; \textbf{(3) suffering from response lag}~\cite{hiebel2019machine, eris2022puppeteer}: These works use features collected in $phase_i$ to infer policies in $phase_{i+1}$. However, the execution state in $phase_i$ may not reflect the memory patterns in the subsequent program execution.}

\subsubsection{\revv{Profile-Guided Software Prefetching}}

\revv{Software prefetching, such as Ayers et al.~\cite{ayers2020classifying}, is limited in access pattern coverage and prefetch timeliness.
It prefetches target addresses by pre-executing prefetch kernels.
For access patterns involving dependent loads, the kernels must first fetch the data from earlier loads, such as regular strided loads, before computing dependent addresses.
This dependency limits how far ahead software prefetching can predict addresses and delays prefetching for subsequent loads.
In contrast, GPS combines multiple hardware sub-prefetchers to cover more diverse access patterns and adjusts their prefetch degrees to improve prefetch timeliness.
}

\subsubsection{\revv{Profile-Guided Temporal Prefetching}}

\revv{Prophet~\cite{li2025profile} uses hints generated from profiling to guide hardware prefetching. However, its hints mainly guide metadata management in a temporal prefetcher, while GPS hints guide the prefetching policies of a composite prefetcher.}

\subsection{\revv{Runtime Data-driven Prefetching}}
\revv{Pythia~\cite{bera2021pythia} uses online reinforcement learning to predict prefetch offsets from a fixed action set rather than coordinate multiple prefetchers.
Like Berti~\cite{navarro-torres_berti_2022}, Pythia~\cite{bera2021pythia} is an individual prefetcher rather than a composite prefetcher policy selector like GPS.
GPS can incorporate Pythia as a sub-prefetcher and coordinate it with other prefetchers.
Figure~\ref{fig:prefetcher_comparison_relative} also evaluates Pythia, whose normalized IPC is 15.08\% below GPS.}

\section{Conclusion}

In this paper, we propose GPS, a data-driven, profile-guided framework for prefetcher selection.
To overcome the limitations of feature misleading, simplistic hardware mapping algorithms, and reactive response lag, GPS shifts the complex policy search to an offline ML model.
By profiling a diverse set of features and applying the IPC-best prefetching policy at each phase, GPS eliminates online exploration overhead.
Evaluation demonstrates that GPS improves geometric-mean IPC over the state-of-the-art I-POP by 12.14\%, 8.34\%, and 14.17\% on SPEC06, SPEC17, and GAP, respectively.

\bibliographystyle{IEEEtranS}
\bibliography{sample-base}

\appendix[AI use]
We used ChatGPT-6 (OpenAI) to assist with English polishing and LaTeX formatting.

\end{document}